\documentclass[10pt, preprint]{aastex}
\usepackage{natbib}

\newcommand{\kms}{km/s}

\begin{document}

\title{High Resolution Spectroscopy of the Pulsating White Dwarf G29-38}
\author{Susan E. Thompson,\altaffilmark{1} J. C. Clemens,\altaffilmark{1,2}
M. H. van Kerkwijk,\altaffilmark{3,4} and  D. Koester \altaffilmark{5}}

\email{sthomp@physics.unc.edu, clemens@physics.unc.edu, mhvk@astro.utoronto.ca, koester@astrophysik.uni-kiel.de}

\altaffiltext{1}{Department of Physics and Astronomy, University of North Carolina, Chapel Hill, NC 27599-3255, USA}
\altaffiltext{2}{Alfred P. Sloan Research Fellow}
\altaffiltext{3}{Astronomical Institute, Utrecht University, PO Box 8000, 3508 TA Utrecht, Netherlands}
\altaffiltext{4}{Present address: Department of Astronomy and Astrophysics, University of Toronto, 
60 St. George Street, Toronto, ON, M5S 3H8, Canada}
\altaffiltext{5}{Institut f\"{u}r Theoretische Physik und Astrophysik, Universit\"{a}t Kiel, 24098 Kiel, Germany}

\begin{abstract}
We present the analysis of time-resolved, high resolution spectra
of the cool white dwarf pulsator, G~29-38.  From measuring 
the Doppler shifts of the H$\alpha$ core, we detect velocity changes
as large as 16.5 \kms\ and conclude that they are due
to the horizontal motions associated with the g-mode pulsations on the star. 
We detect seven pulsation modes from the velocity time-series and
identify the same modes in the flux variations.
We discuss the properties of these modes and use the advantage
of having both velocity and flux measurements of the pulsations to test the 
convective driving theory proposed for DAV stars.  
Our data show limited agreement with the expected relationships between the amplitudes 
and phases of the velocity and flux modes. 
Unexpectedly, the \emph{velocity} curve shows evidence for harmonic
distortion, in the form of a peak in the Fourier transform whose frequency
is the exact sum of the two largest frequencies.  Combination frequencies
are a characteristic feature of the Fourier transforms of light curves of
G~29-38, but before now have not been detected in the velocities, nor does
published theory predict that they should exist.  We compare our velocity
combination frequency to combination frequencies found in the analysis of
light curves of G~29-38, and discuss what might account for the existence of
velocity combinations with the properties we observe.

We also use our high-resolution spectra to determine if either rotation or pulsation
can explain the truncated shape observed for the DAV star's line core. We 
are able to eliminate both mechanisms:
the average spectrum does not fit the rotationally broadened model 
and the time-series of spectra provides proof that the 
pulsations do not significantly truncate the line. 

\end{abstract}


\section{Introduction}
G~29-38 (ZZ Psc) is an extensively observed V=13.05 magnitude pulsating
white dwarf that lies at the cool end of the DA instability strip.
Like many cool DAV stars, its oscillation properties are highly variable.  In one month it
may exhibit a dominant period of 615~s and amplitudes of 6\% \citep{Wi90}.  
In another month it
may have a dominant period of 809~s and amplitudes of 4\% \citep{K98}. 
Occasionally it
shows no measurable pulsations at all.  In addition to being confusing in
themselves, these amplitude changes frustrate attempts to identify
eigenfrequencies that might allow seismological analysis of the internal
structure of G~29-38. 

   \citet{K98} accomplished a breakthrough in seismology of G~29-38
by compiling and analyzing many months of time-series photometric
observations.  They found a consistent pattern of recurring frequencies in the
data. This result is reassuring to seismologists because it suggests that the
observed frequencies represent eigenfrequencies that carry information about
internal structure. \citet{K98} and \citet{BK97} presented an analysis of the pattern
of modes, concluding they are most likely modes of the first spherical harmonic
degree ($\ell =1$). The conclusion was based on the match of the spacing between modes to
theoretical expectations for $\ell =1$ modes of consecutive radial eigennumber, and
upon the observation of triplet structure in some of the modes.  Rotation should
split $\ell =1$ modes into three components of different azimuthal quantum number ($m$).
Seismological analysis of the mode pattern yielded a
mass estimate of $0.6M_\sun$ for an assumed Hydrogen layer mass of $10^{-4}M_\star$.

\citet{K98} had no way to verify their mode identification by
direct observation, but subsequent time-series spectroscopy at high
signal-to-noise by Clemens, van Kerkwijk, \& Wu (2000) provided verification for some modes.
Because of the wavelength dependence of limb darkening, pulsation amplitudes at each
color are sensitive to the degree ($\ell$) of pulsation modes.  Using time-resolved
spectroscopy from Keck II LRIS \citep{Oke95}, \citet{C00}
confirmed that most of the modes are $\ell=1$.  The only exception is a mode
at 777~s that \citet{C00} identify as $\ell=2$. 

In addition to mode identification, 
van Kerkwijk, Clemens, \& Wu (2000) 
detected velocity shifts associated with pulsations of G~29-38, and thereby obtained the first
direct measurements of the size of the surface motions in a pulsating white dwarf.  Because
surface velocities are distributed as the derivative of a spherical harmonic,
their integrated contribution does not cancel in the same way as the spherical
harmonically distributed flux variations.  As a result, the ratio of velocity to
flux amplitudes is sensitive to $\ell$, and measurements of these ratios also support
the conclusion that the 777~s mode is $\ell=2$ and the remainder $\ell=1$.  Later,
Kotak, van Kerkwijk, \& Clemens (2002) showed that the 918~s mode observed 
by \citet{VK2000} also has $\ell=2$.

Encouraged by the success of the LRIS observations \citep{VK2000}, we aspired to 
improve the velocity measurements by observing at higher resolution. 
Accordingly, we obtained a 
5 hour time-series of the H$\alpha$ core of G~29-38 with the Keck I High 
Resolution Echelle Spectrograph \citep[HIRES]{Vo94}. We used a narrow slit to improve the 
velocity measurements of G~29-38 at the expense of some quality in the flux measurements.
Using this technique, we have sucessfully detected the pulsation velocities, 
confirming the LRIS observations of \citet{VK2000}, and we have acheived 
a signal-to-noise in the velocity curve better than any previous measurements.
In this paper we use our higher precision measurements to examine questions 
concerning the ZZ Ceti driving mechanisms, combination modes, and the shape 
of the ZZ Ceti H$\alpha$ line.

Recent theories of ZZ Ceti mode excitation make specific predictions about 
the relative sizes of velocity and flux variations. Our simultaneous 
measurements of each allow us to test the predictions. Originally, the 
classical $\kappa-\gamma$ mechanism was proposed as a plausible driving mechanism for the pulsations 
\citep{Wi82, DK81, DV81}.  Later, \citet{Br83, Br91} recognized that the convection zone responds
immediately to the flux perturbations from the radiative interior and demonstrated 
that the convection zone could be the center of mode driving.  
His convective driving mechanism, as further developed by \citet{GW99}, 
makes explicit, testable predictions concerning the relationship between the 
phases and amplitudes of the observed modes.
Our measurements agree with the basic predictions of this model, but are not sufficient to
provide rigorous tests of how mode qualities change with frequency.
  
Our observations also enable us to explore the
harmonic distortions frequently observed in G~29-38.  
The distinct shape of ZZ Ceti light curves 
have been a puzzle since \citet{WN70} published the first
time-series photometry of HL Tau 76.  Small-amplitude ZZ Ceti pulsators,
now identified as the hotter members of the class, have nearly sine-like
photometric variations, but the cooler, large-amplitude pulsators show
distinctive non-sinusoidal shapes, with broad, flat minima and sharp maxima
(see Figure 1).  Corresponding to these shapes are the appearance of sum and
difference frequencies in the Fourier spectrum.

\begin{figure}
\figurenum{1}
\plotone{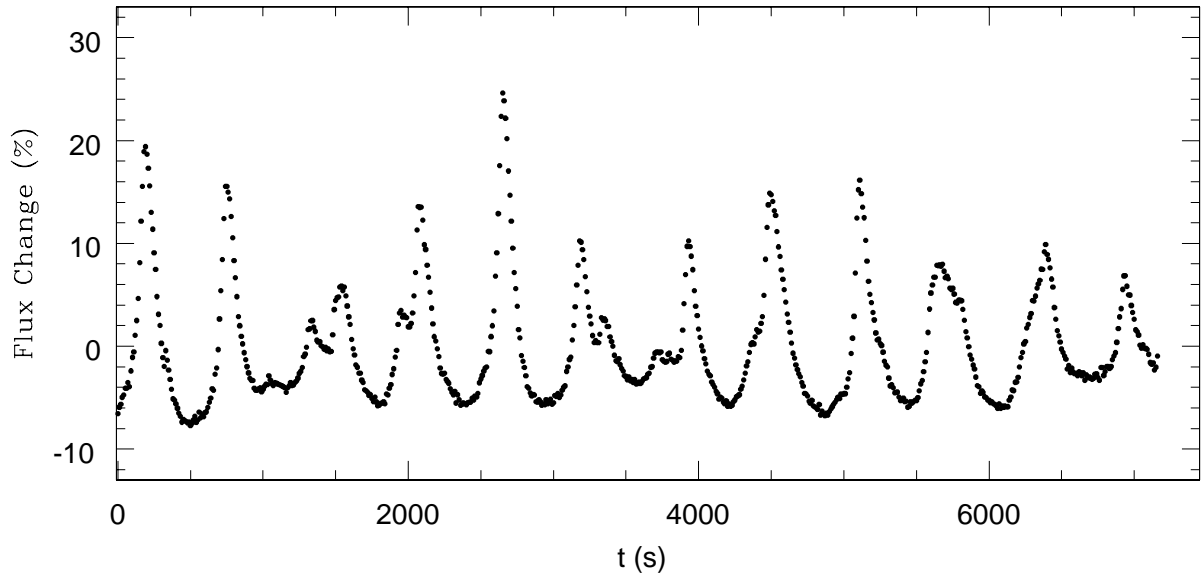}
\caption{A typical light curve of G~29-38 showing the non-sinusoidal character of the pulsations, 
courtesy of the Whole Earth Telescope data
archives. These data were obtained by M. Wood using
the 82" Struve telescope at McDonald Observatory and a Nather two-star photometer
with 10 s sampling times.}
\end{figure}

Theoretical calculations, first by \citet{Br83, Br92}, and later by 
\citet{Wu01} and \citet{IK01}, expain this phenomenom within the context
of the convective driving theory.  Brickhill
built a numerical model of a ZZ Ceti surface convection zone and showed that
sinusoidal flux variations at its base are attenuated and delayed in phase
before reaching the surface.  Moreover, because the thickness of the model
convection zone decreases between flux minimum and flux maximum, the amount
of attenuation and delay changes within a cycle.  This non-linearity
distorts the sinusoidal input variation into a surface flux variation whose
shape in the model is similar to that we observe.  \citet{Wu01} has found the
same result analytically, while \citet{IK01} have confirmed and
extended the numerical results of \citet{Br83}.

  Other theories for the origin of the pulse shapes invoke the non-linear
relationship between flux and surface temperature \citep[$F \propto T^4$;
see][]{Br95}, or the excitation of independent pulsation modes by
resonant mode coupling.  In a recent series of papers, Vuille and
collaborators show that the former theory cannot account for the large sizes
of the combination frequencies in the large amplitude pulsator G~29-38
\citep{VB00} and the latter cannot naturally explain the
phasing of modes with their combinations \citep{Vu00b}.  They conclude that the
majority of combination frequencies in G~29-38 and, by inference, the other
large amplitude pulsators arise from harmonic distortion like that described
in the models of \citet{Br83, Br92}, \citet{Wu01}, and \citet{IK01}.

Our measurements complicate matters further by 
revealing the first combination frequency in the Fourier transform of the \emph{velocity}
curve of a ZZ Ceti star. This detection is a surprise, because
\citet{Br83} and \citet{Wu01} both maintain that surface velocities should not be
distorted by the convection zone in their models.  We will speculate
about what might account for our observations, using the relative phase of the velocity 
combination we have measured as our only clue.  Because we have detected only a 
single combination peak, we will require more observations to establish whether 
or not its properties are typical of large amplitude ZZ Ceti pulsations.

Finally, our spectra of G~29-38 help us explore another mystery, 
associated with the shape of the H$\alpha$ core in the mean spectrum of DAV stars.
\citet{Ko98} observed the NLTE H$\alpha$ core
profiles for 28 DA white dwarfs in order to measure their rotation rates.   For the
majority of DA stars, they measured line cores consistent with $v\sin i$ of 15 \kms\
or less, but for all 3 ZZ Ceti stars in their sample (including G~29-38) and one 
star just below the instability strip, they found shallow line cores 
requiring velocities of 29-45 \kms.  
Not only are rotation speeds of this size too high for the rotation period suggested
by triplet splitting in the \citet{K98} study, but they require that
the ZZ Cetis as a group rotate much faster than other DA stars.  This would indicate
that ZZ Cetis are either not ordinary DAs or their rotation rate increases when
they cool into the instability strip.  Our mean spectrum confirms the measurements
of \citet{Ko98}, and we explore whether the 
Doppler shifts associated with the pulsations could cause the unusual line 
shape for G~29-38. 
 
The paper is organized in seven sections. In the next section (\S 2) we 
describe the observations and the 
method used to reduce the spectral images into velocity and flux curves.  
In \S 3 and \S 4 we examine the modes revealed via Fourier analysis 
and discuss the results in the context of the convective driving model.
In \S 5 we discuss the combination frequency discovered in the velocity
curve.  In \S 6 we compare our average spectrum to that
of \citet{Ko98} and discuss the cause of the anamolous line core. 
Finally, we present our conclusions in \S 7.

\section{Observations and Reduction}

On September 16, 1997, we acquired a series of 236 spectra of G~29-38 using 
HIRES \citep{Vo94} on the Keck I telescope.
We used the red collimator, the KV408 order blocking filter, and the
C2 decker, which yielded a slit width of 0.861''. In this configuration,
the 24-micron pixel Tektronix 2048EB2 CCD with 2x2 binning gave a dispersion of
0.095 \AA\ per binned pixel, and a scale of 0.38" per binned pixel in the
spatial direction.  The effective spectral resolution, defined by the slit width,
was 0.215\AA.  

We began our 5 hour sequence of exposures at 08:28:02 UT (measured by the
observatory clock and recorded in the image headers). The exposure time for each
spectrum was 50 seconds with an additional 28 seconds required to read and wipe
the CCD, yielding a duty cycle of 64\%.  Since reading the entire CCD would
have taken much longer and thus unacceptably reduced the duty cycle, we only read a
region containing three orders (53, 54, and 55) with the middle order centered on
H$\alpha$.  For this small region, the readout time with two amplifiers was not
significantly better than with one, so we used one amplifier to eliminate
complications arising from differences in gain and linearity.  The exposure
levels in our 50 second integrations ranged from 35 electrons in the core of
H$\alpha$ to 70 in the wings.  The moon was full, so the background sky was
relatively high, ranging from 16 to 40 electrons with clearly visible features of
the solar spectrum, and the shot noise from the sky is comparable to the read noise of
4.5 electrons. In Figure 2 we show a raw spectral image for
the average of all 236 images. We reduced only the central order
since we were interested in the H$\alpha$ line and did not need the extended 
continuum in the other orders.

\begin{figure}
\figurenum{2}
\plotone{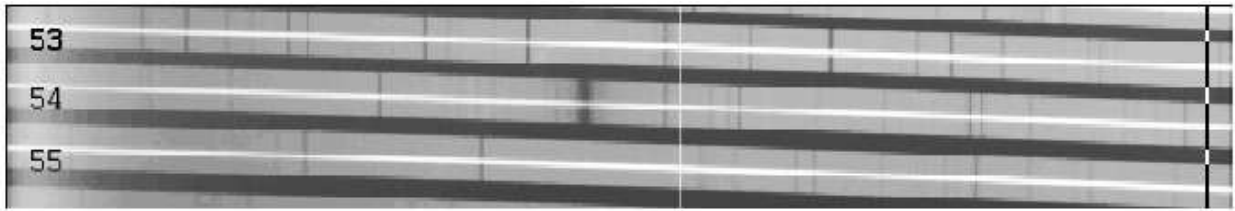}
\caption{The average of the spectral images prior to reduction. The echelle orders are labeled on the left. H$\alpha$ is located at the center of the 54th order.}
\end{figure}

We acquired well-exposed spectra of a halogen lamp to use
for spectral flats and of a Thorium-Argon lamp to use for wavelength
calibration.  We also acquired spectra of the flux standard G~126-18, but in this
paper we present only relative photometry, and have made no attempt at absolute
flux calibration.  

The seeing during our run, as measured from spatial profiles,
averaged about 1.2" fwhm, so the relatively narrow slit makes photometry
problematic.  Moreover, to avoid introducing potential periodic variations, we
did not use the image rotator; so differential refraction slowly changed the
amount of light passing through the slit.  As we shall see, these losses did not
prevent us from measuring periodic variations in the brightness of G~29-38.

We reduced our 236 spectral images using standard Image Reduction and
Analysis Facility (IRAF) routines \citep{To86}.  We removed the bias level by fitting and
subtracting a plane to the prescan region of each image.  We removed the blaze
function and the flat field variations in one step by dividing the data by the
normalized average of the halogen lamp exposures.  Normalization was accomplished
by dividing the average image by the mean pixel value of the image. To extract individual spectra,
we first calibrated our extraction parameters using the average spectrum (see
Figure 3) and the IRAF apall routine.  This gave values for the slope, curvature,
and spatial profile of the average spectrum.  We assumed individual spectra would
have similar values, but might be shifted along the slit, and conducted optimal
extraction \citep{Ho86} of the lower signal-to-noise, individual spectra with spatial position
as the only free parameter. The measured motion of G~29-38 along the slit provides an
estimate of the motion across the slit; we used these to assess the slit losses 
and wavelength stability.

Figure 3 shows a typical individual spectrum after reduction and
wavelength calibration.  The signal-to-noise is about 8 in the wings of the line
and 5 in the line core.  The lower panel shows an average of all 236 spectra and
was constructed by averaging the reduced spectral images and performing a single
extraction.  We extracted two time-series from the individual spectra, one
representing the changes in velocity and the other representing changes in flux. 

\begin{figure}
\figurenum{3}
\plotone{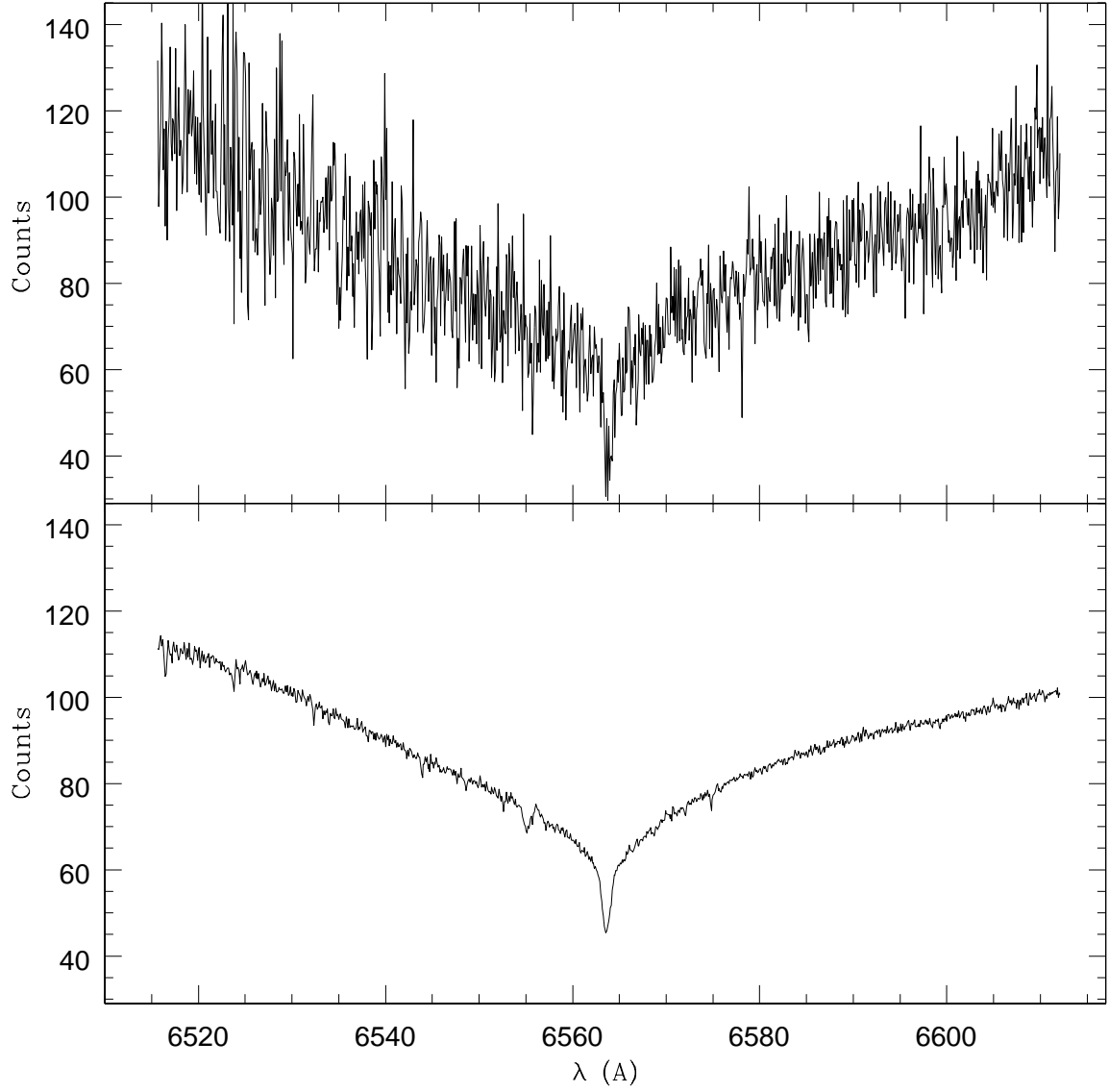}
\caption{An example of an individual spectrum (top panel) and the spectrum extracted from the average of all 236 spectral images (bottom panel). The dominant feature at $~$6555 \AA\ is due to scattered light. Other features are telluric water lines.}
\end{figure}

To examine the line-of-sight velocity modes on the star, we measured the Doppler
shift of the H$\alpha$ line in each spectrum.  Attempts at using cross-correlation
against a smoothed average spectrum yielded inconsistent results; 
the amplitude of the shift was skewed by the amount of the 
spectrum included in the correlation.
As an alternative, we fitted each spectral line to obtain the location of the central
wavelength.  A Gaussian and Lorentzian, forced to have the same central
wavelength, super-imposed on a linear continuum, were fit to the average spectrum
over the range 6557.5-6570.0\AA.   Using this description of the line, we fit the
average spectrum until no significant improvement in chi-square was achieved.
We obtained a reduced chi-square of about 0.5, indicating a good fit.  We then fit each individual spectrum using
the average spectrum's fit as the initial conditions; allowing only the central wavelength, 
the height of the continuum, the slope of the continuum, 
and the flux contained in both the Gaussian and Lorentzian to vary for each fit. The width 
of the Gaussian and the Lorentzian were fixed to the average spectrum's fitted values.  
A subtraction of the central wavelength, established with the average spectrum, yielded the spectral
shift for each spectrum, and thus line-of-sight velocities.  We observe Doppler shifts as large 
as 0.36\AA\ (3.8 binned pixels), corresponding to a velocity of 16.5 \kms. Figure 4a shows 
the velocity curve resulting from this reduction.

To obtain the relative changes in the flux of the star, we summed each spectrum
from 6553 to 6573 \AA\ and divided the resulting light curve by the mean of
these values.  The light curve, presented in the lower panel of Figure 4, is
normalized by a second order polynomial fit in order to remove long period variations introduced
by extinction and differential refraction and yield fractional changes in flux. 
We observe flux variations as large as 20\% from the  
mean flux. Two sections of the light curve, as indicated 
in Figure 4, appear to be obstructed by clouds and thus not 
included in the analysis of the light curve. Increased sky levels concurrent with the decrease in star 
light confirms this conclusion for the latter region.
The Fourier transform and least squares fits of the light curve in \S 3 do not include these regions.  

\begin{figure}
\figurenum{4}
\plotone{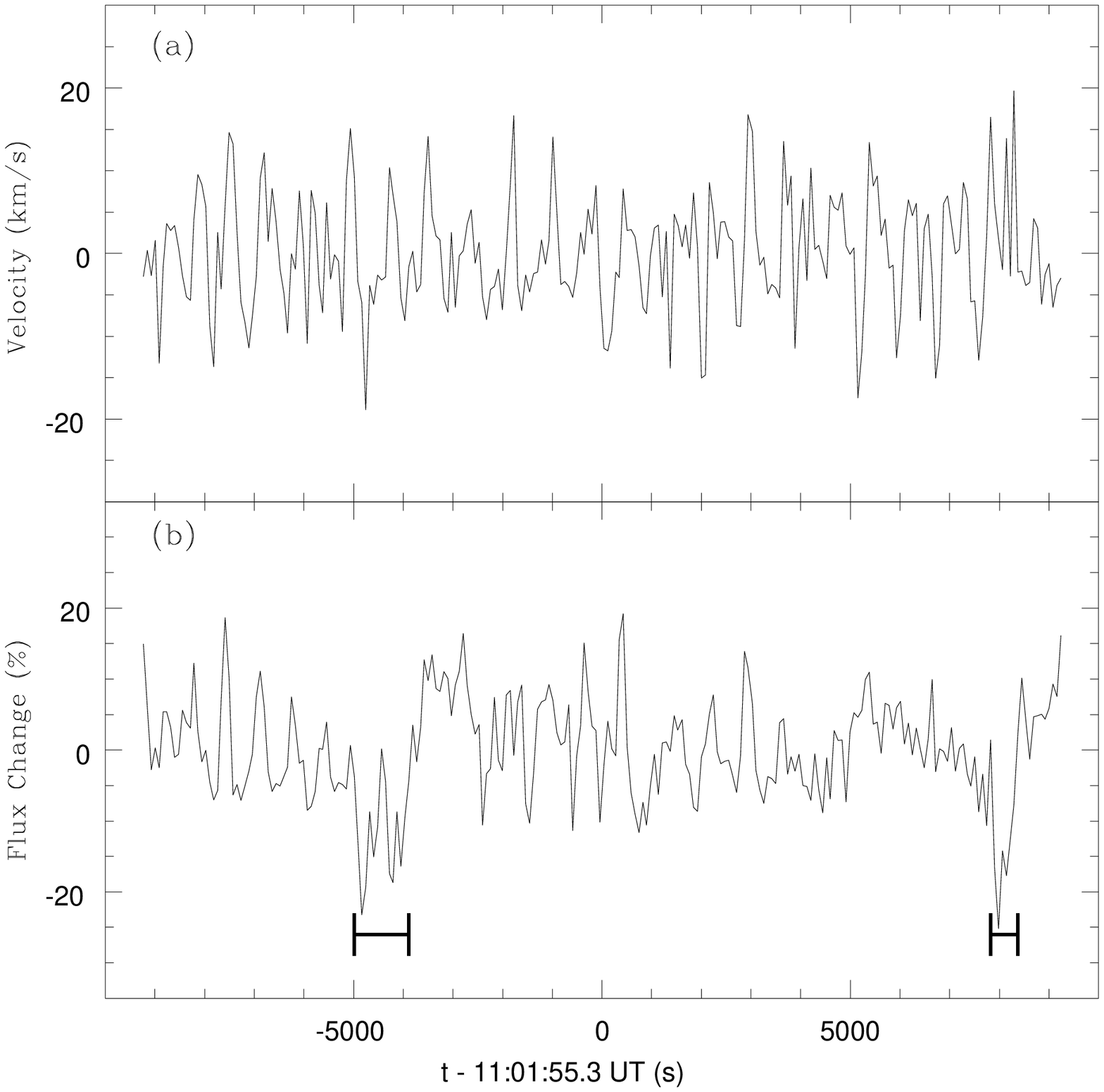}
\caption{\textbf{(a)} The line-of-sight velocity curve, where a positive velocity indicates a red shift. A typical error on a velocity measurement is 3 \kms. \textbf{(b)} The change in flux normalized by a fitted parabola. The marked regions are not used in the Fourier transform or the least-squares fits of the flux curve.}
\end{figure}

We measured the effect of flexure on our data by performing this
same fitting technique on the spectrum of the sky background in each spectrum.
The central wavelength of the sky showed little trend and only varied by at most
0.2 pixels ($\sim 1$~\kms).  

To be certain the velocities detected are not due to the star moving
perpendicular to the slit, we estimated the loss of light that would also result 
from this motion.  
We fitted a Gaussian distribution to the spatial profile and assumed the same profile 
in the dispersion direction (correcting for the different spatial scales). 
For a 0.36 \AA\ shift (16.5 \kms), the largest 
detected spectral shifts, we would expect to see flux variations as large as 30\%
correlated with the shifts. 
The largest measured flux variations are less than 20\% and 
are not concurrent with the largest spectral shifts. Furthermore, 
if stellar wander caused the observed spectral shifts, we would expect
to see peaks at half the period of the velocity modes. Looking
ahead to the Fourier transforms of the velocity and flux variations (Figure 5), we
do not see peaks of this sort. Thus, we conclude that stellar motion in the 
slit is not the primary source of the detected wavelength shifts. 

\section{Flux and Velocity Periodicities}

We began analysis of the velocity and flux time-series by investigating 
the periodicities present in both.  Figure 5 shows
the Fourier transforms (FTs) of the velocity and flux curves introduced 
in \S 2. The three dominant modes
in both curves appear at approximately the same frequencies, indicating that the
velocity and flux modes are correlated.  As such,
our data confirm the results of \citet{VK2000}, showing significant
line-of-sight velocity variations due to the g-mode pulsations on the star.

To further convince ourselves that these Doppler shifts exist in the spectra, we
averaged together individual spectra with similar velocities. Figure 6 shows two
spectra created by averaging together 30 spectra fit with large positive
velocities (red-shifts greater than 6.9 \kms) and 30 spectra with large negative
velocities (blue-shifts greater than 6.9 \kms).  A noticeable wavelength shift in 
the center of the line can be seen between these two 
averages. Hence, we conclude that these Doppler shifts 
are indeed apparent in the data and not just artifacts of the fits.
  
\begin{figure}
\figurenum{5}
\plotone{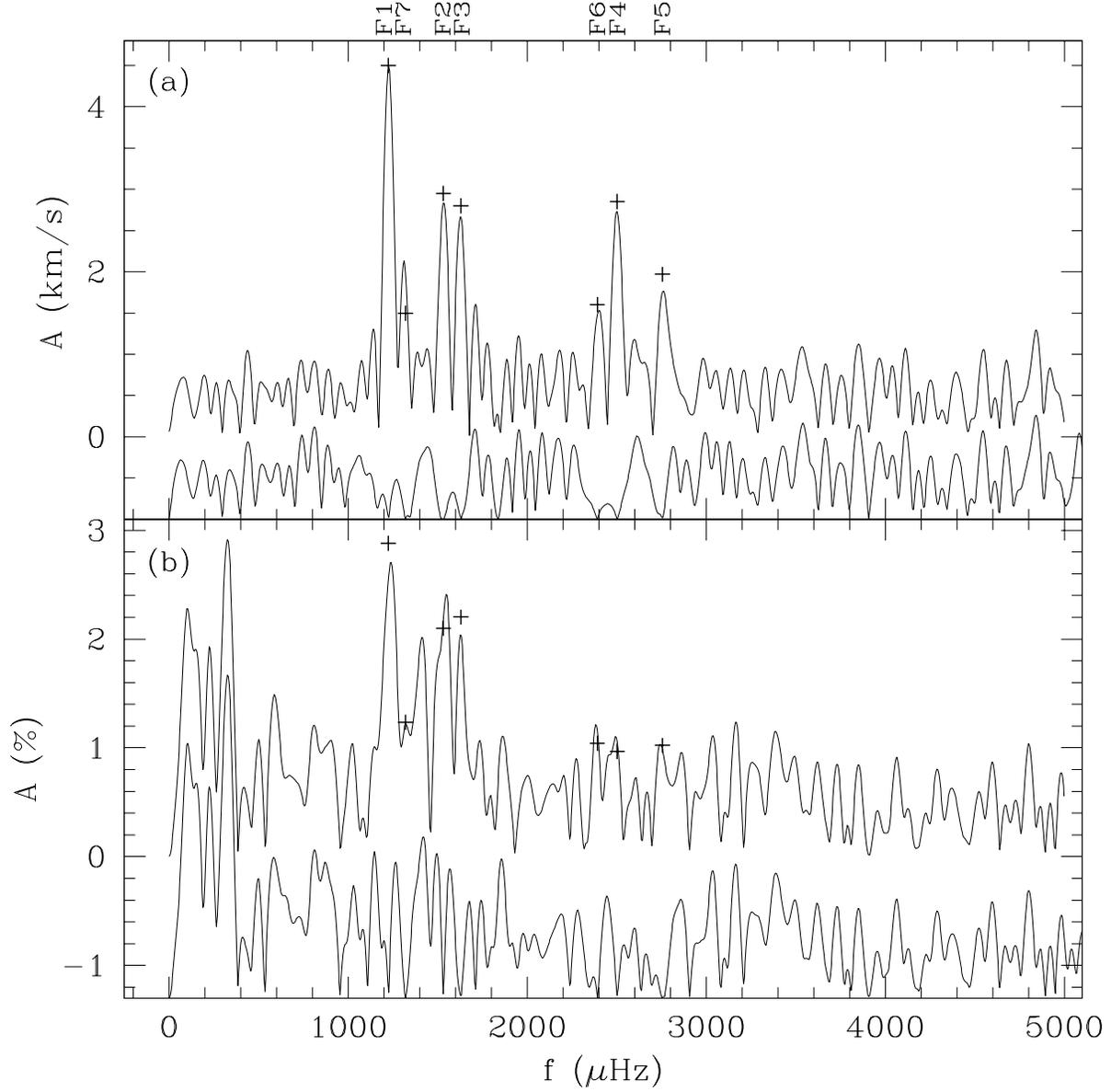}
\caption{The amplitude Fourier transforms of the line-of sight velocity curve \textbf{(a)} and relative flux curve \textbf{(b)}.  The seven modes are marked at the extracted values in both FTs. The amplitude FTs of the residuals are plotted below each FT. The flux residual is off-set by 1.3\% while the velocity residual is off-set by 1.0~\kms. The low frequency peaks not extracted from the flux transform are probably atmospheric effects and not inherent to the star.}
\end{figure}

The noise level in our velocity curve is significantly better
than that measured by \citet{VK2000} while the noise in the flux is
worse.  This is expected since our technique used a narrow slit to improve
velocity measurements at the expense of flux. Our lower noise velocity 
curve makes the detection and identification of frequencies more reliable 
relative to our flux measurements.  Moreover, there are significant peaks 
in the velocity transform with no apparent
counterparts in the flux transform.  Consequently, we analyzed the velocities
first and then fitted the flux amplitudes under the assumption that the modes
identified have the frequencies measured in the velocity transform.

We began by identifying the periodicities in the velocity curve and 
removing them consecutively in order of decreasing
amplitude.  Using the FT's value for the frequencies and phases as the initial
conditions, the velocity curve was fit to a function of the form
$A_v\cos (2\pi ft-\Phi_v )+C$ where $A_v$ is the amplitude, $f$ is the frequency,
$\Phi _v $ is the phase and $C$ is a constant offset.  We removed this 
fitted cosine wave from the data and then repeated the process by identifying the 
next frequency until no significant peaks could be determined from the Fourier spectrum. 
A final simultaneous fit of the data to 
a linear combination of the cosine waves for all identified frequencies 
yielded the values for $A_v$, $f$ and $\Phi_v$ (Table 1). We used the average of the time 
stamps on the 236 spectra (11:01:55.3 UT) as the zero point for determining the phase.

We estimated the expected noise in the velocity time-series by considering two
components: one from the wandering of the star's position within the
slit and the other from our ability to fit the line center in the relatively noisy
individual spectra. By measuring the motions of the star along the slit, and removing a
parabola to account for slow changes in position with airmass, we constructed an
estimate of the rms amplitude of the motions.  If we assume that motions across
the slit have the same amplitude, this translates into an rms velocity error of
0.4 \kms\ with the largest extent being 1 \kms\ (0.36 pixels spatially).  
To assess the possible impact of periodic wandering, we calculated a Fourier transform 
of the motions. It shows significant peaks at various
guiding frequencies, but they are all far smaller (a factor of 10 to 20 times)
than the velocity amplitudes detected in G~29-38 and do not have the same
frequencies.  The formal error of our velocity fits to individual spectra is $\sim 3.8$
\kms, much larger than the errors from slit motions.  If we assume these are
Gaussian, then they should translate into an average noise in the
fourier transform of 0.5 \kms. Using this noise level, our fits to 
the velocity curve yield a reduced chi-square of about 1.4, 
confirming that 0.5 \kms\ is a slight underestimate of the noise. 
We can get a separate estimate of the noise by averaging the 
power in the 3.5-5.0~mHz region of 
the velocity FT. This yields an error of $\sim 0.6$ \kms. 
We believe that the slight excess may be intrinsic variations 
due to the low-amplitude pulsations, which are known to exist in this frequency range.
The noise level of 0.6 \kms\ gives a reduced chi-square of about 1.0; we
use this noise level to determine the errors reported in Table 1.

\begin{deluxetable}{rcccccccc}
\tablenum{1}
\tablecolumns{9}
\tabletypesize{\footnotesize}
\tablewidth{0pt}
\tablecaption{The amplitudes, phases, $R_v$, and $\Delta\Phi$ for each mode.}
\tablehead{\colhead{mode} & \colhead{P(s)} & \colhead{$f$($\mu$Hz)} & \colhead{$A_v$(km/s)} & \colhead{$\Phi _v$ (\degr )} & \colhead{$A_L$ (\% )} & \colhead{$\Phi _L$ (\degr )} & \colhead{$R_v$ (Mm/rad)\tablenotemark{\dag}} & \colhead{$\Delta \Phi$ (\degr ) }}
\startdata
F1&816&$1224.9\pm 3.0$&$4.53\pm 0.44$&$270\pm 6$&$2.88\pm 0.55$&$224\pm 11$&$20.4\pm 4.4$&$46\pm 12$\\
F2&653&$1532.1\pm 4.6$&$2.95\pm 0.44$&$216\pm 9$&$2.11\pm 0.54$&$165\pm 15$&$14.5\pm 4.3$&$52\pm 17$\\
F3&614&$1629.1\pm 4.8$&$2.79\pm 0.44$&$304\pm 9$&$2.20\pm 0.54$&$253\pm 14$&$12.4\pm 3.6$&$50\pm 17$\\
F4&399&$2503.4\pm 4.7$&$2.84\pm 0.44$&$177\pm 9$&$0.97\pm 0.54$&$12\pm 32$&$18.7\pm 10.8$&$165\pm 33$\\
F5\tablenotemark{*}&363&$2754.9\pm 6.7$&$1.97\pm 0.43$&$31\pm 13$&$1.02\pm 0.54$&$35\pm 30$&$11.1\pm 6.4$&$-5\pm 33$\\
F6&418&$2391.9\pm 8.5$&$1.57\pm 0.44$&$280\pm 16$&$1.04\pm 0.54$&$4\pm 30$&$10.0\pm 5.9$&$275\pm 34$\\
F7&757&$1321.1\pm 9.3$&$1.48\pm 0.45$&$137\pm 17$&$1.23\pm 0.54$&$260\pm 25$&$14.4\pm 7.7$&$-124\pm 31$\\
\enddata
\tablenotetext{*}{F5 is the combination mode, its frequency is the sum of F1 and F2.}
\tablenotetext{\dag}{The reader may prefer the more intuitive units of $10km/rad/\%$.}
\tablecomments{For each mode, $R_v = (A_v/\omega)/A_L$ and $\Delta\Phi = \Phi_v -\Phi_L$. 
The reported errors reflect a reduced chi-squared of 1 for the least-squares fits 
to the flux and velocity curves.}
\end{deluxetable}

To determine the significance of the identified velocity peaks, we conservatively
used the noise level determined from the background average power of the FT (0.6 \kms). 
As such, all 7 modes lie at
or above $3\sigma$.  However, to be certain that these amplitudes could not occur by chance, 
we calculated the false alarm probability \citep{HB86,Ke93}. The five largest velocity 
modes have less than 1\% chance of only being due to noise.  The significance
of F6 and F7 are questionable since they  
have a false alarm probability of $\sim \onehalf$. They appear more significant in 
the FT because the side lobes of the F1 and F4 modes respectively increase their apparent size.  

After completing our fits to the velocity curve, we extracted the 
periodicities in the flux curve by fitting it with 
a combination of cosine waves, fixing the frequencies to the fitted 
velocity frequencies. Table 1 contains the amplitudes and phases for the
seven modes extracted using this technique.
The low frequency peaks in the flux FT are not included since they are not present in the velocity
curves and likely result from atmospheric effects, and from  
an observational technique not optimized for flux
measurements. Consequently, the flux curve is not well accounted for by the linear
combination of the seven cosine waves. To obtain the errors from the fit we scaled up the 
error on each measurement to obtain a reduced chi-square of 1. This is reflected in 
the errors presented in Table 1.    

Since we measured the flux modes by fixing the frequencies to the 
velocity mode frequencies, we determined their significance by calculating
the odds that noise would create a peak of that amplitude at the exact frequency 
we expected to see one.  We estimated the noise level from the average of 
the flux power FT in the region from 3.5-5.0~mHz to be $0.5 \%$. 
The smallest flux peaks (F4, F5, and F6) lie around $2.3\sigma $ and have a 
2\% probability that Gaussian distributed noise would show a peak 
of that power at the expected frequency. 

The frequencies we have measured are all similar or identical to g-mode
frequencies previously measured in flux variations of this star.  The mode
F4 is almost, but not exactly, two times the largest mode F1.  Both of these
modes are listed in the tabulated photometric modes for G~29-38
\citep[see][]{K98, Vu00a}
Near-resonances of this sort are common in the large
amplitude ZZ Ceti stars \citep[e.g. BPM 31595;][]{OD92} but not understood.
They do not
originate from harmonic distortion of pulse shapes, because that mechanism
generates exact frequency combinations.

However, the velocity signal at F5 is the exact sum of the two largest
frequency modes (F1+F2) to within the measurement error, as we might expect from pulse shape
distortion.  We do not see any other combinations or harmonics in the
spectrum.  It is quite common in flux measurements to see a signal
at the sum of two large modes, but not be able to detect their individual
harmonics.
In this respect, the velocity transform of Figure 5 resembles a typical flux
transform for G~29-38.  However, the existence of a velocity combination peak presents a
challenge for the convective driving theory, which does not predict distortion in velocity variations. 
We discuss a possibility for answering this challenge in \S 5 below.

Because F5 is unexpected, we have worried about possible non-linearities in
our fitting procedure that might generate a false combination signal.  If
the spectral fitting process is non-linear in some way, or if it is biased
by the flux, which is varying at the same frequencies but different phase,
then harmonics or combination frequencies might be generated as artifacts
during the analysis. We have experimented with different fitting routines
and also applied cross correlation between the average spectrum and each
individual spectrum.  In every case, the time series shows a signal at F5,
although its significance is sometimes lower.  Furthermore, none of the
techniques generates combinations or harmonics of other modes in the Fourier
transform.  We conclude that F5 is a combination frequency present in the
star, but emphasize the need for further observations to establish whether
velocity combinations are a persistent feature of large amplitude ZZ Ceti
pulsators.  The first measurement of pulsational velocities by \citet{VK2000}
did not show velocity combinations, but the signal-to-noise in their
velocity curve
was less than one-half of ours, and they would not have been able to detect
a signal the size of our F5.

\section{Testing the Convective Driving Theory}  

With the amplitudes and phases of the seven detected pulsation modes, 
we have knowledge of the motions and the temperature changes 
at the surface of this ZZ Ceti.
The usefulness of these measurements goes beyond calibrating
the size of the velocity variations at the surface of G~29-38;
it allows us to test specific predictions of the convective driving theory,
concerning the relationship between the flux and velocity of the modes. 
In the convective driving mechanism described by \citet{Br83} and \citet{GW99},
driving occurs because of the convection zone's response to the
flux perturbations.  From this interaction, the flux variations are 
distorted while the velocities basically traverse the convection zone undiminished. 
Having both velocity and flux measurements 
of the pulsation modes, we are able to test the 
convective driving theory by comparing our data
with the analytic predictions of \citet{GW99} and the 
numerical model of \citet{GW99b}. 

We have presented the frequency, amplitudes and phases of all seven 
extracted modes in Table 1. Of these modes, the first four 
have been previously observed in this star \citep{K98, VK2000}.
To see the relationship between the two measurements of the modes, we calculated 
the relative amplitude, $R_v=(A_v/\omega)/A_L$, and the difference 
between the velocity and flux phases, $\Delta \Phi = \Phi_v -\Phi_L$, 
in accordance with the conventions used by \citet{VK2000}. 
The relative amplitude is a ratio of the velocity amplitude ($A_v$), 
scaled by the radian frequency ($\omega = 2\pi f$), 
and the flux amplitude ($A_L$); by including the frequency, $R_v$ depends
only on $\ell$, and not on frequency, for adiabatic pulsations \citep[see][]{Dz77,C00}. 
Only $\Delta \Phi$ of the first three modes were confidently measured
and have values around $50\degr$. 
The other modes have such low flux amplitudes that the fitted phases 
could have been easily swayed by noise.
Both the relative amplitudes and the phase differences have values
similar to the previous measurements of this star \citep{VK2000}. 
The $R_v$ range from 10 to 20 $Mm/rad$; none is as high as the mode 
identified as $\ell =2$ by \citet{VK2000} and \citet{C00} ($R_v=27\pm 11\ Mm/rad$ for the 777~s mode). 
Hence, these modes likely have a
spherical degree of one, as expected from the conclusions of \citet{K98},
and measured directly for F1, F2 and F3 by \citet{C00}.

We can interpret some of these measurements in the context of
the analytical theory of the convective driving mechanism \citep{GW99}. 
During an adiabatic pulsation, the maximum velocity
occurs a quarter cycle after the flux maximum. However, the large heat capacity of
the convection zone alters this basic picture by bottling up heat, thereby reducing the 
amplitude and delaying the phase of the flux perturbations. This will decrease
the difference between the phases of the velocity and flux maxima and increase the
ratio between the velocity and flux amplitudes.

Our measurements agree (see Table 1) with these basic predictions of the model;
we observe the values of $\Delta\Phi$ to be less than $90\degr$ 
and $R_v$ is larger than expected for an adiabatic mode \citep[see][]{VK2000}.
We can also specifically compare our data to the model since \citet{GW99b} 
calculated their model for a star similar to G~29-38. 
For an 800~s mode on a 12,160~K star 
with $\omega \tau_{th} = 0.8$, 
their model shows a phase difference of 
$55\degr$ and relative amplitude of 16 $Mm/rad$. 
The 818~s mode measured by \citet{VK2000} 
confirmed this prediction ($\Delta \Phi =44\degr\pm 19\degr$; $R_v = 11 \pm 4\ Mm/rad$).
Our measurements of F1 at 817~s mode also agree within the errors to the model calculations.
For a graphical comparison, Figure 7 plots our data along with the results of this model.  

Another result of the convective driving mechanism \citep{GW99}
is the frequency dependence of the interaction between the
convection zone and the temperature variations.  
For larger values of the product of the thermal 
adjustment time, $\tau_c$, and the radian frequency, $\omega$, the convection 
zone has a larger effect on the flux variations. In fact, 
convective driving can only cause overstability of a mode when the 
inequality $\omega \tau_c \geq 1$ is satisfied. As such, we 
are able to determine a low estimate of the thermal
adjustment time from the longest period mode ($\tau_c \geq 130$ s). 

From this $\omega \tau_c$ dependence, we expect frequency trends 
in the relative amplitude and phase difference. 
With increasing frequency, more flux is absorbed, reducing 
the measured flux amplitude ($A_L$) with respect to the distance the 
material travels on the star ($A_v/\omega$). Thus, the relative
amplitude should increase.  Also, with smaller periods,
the flux variations experience more phase delay and the 
difference between the velocity and flux phases
become smaller and farther from $90\degr$. 
\citet{GW99} quantifies these trends with expressions
for the attenuation and phase delay of the 
flux variations. The observed flux amplitude is attenuated 
by the factor $(1+(\omega \tau _c)^2 )^{-1/2}$ and
the flux phase is delayed by $\arctan (\omega \tau_c)$.
Thus, because of the presence of $\omega \tau_c$ in these expressions, 
larger frequencies in the same star should have larger values of 
$R_v$ and smaller values of $\Delta \Phi$.

\begin {figure}
\figurenum{6}
\plotone{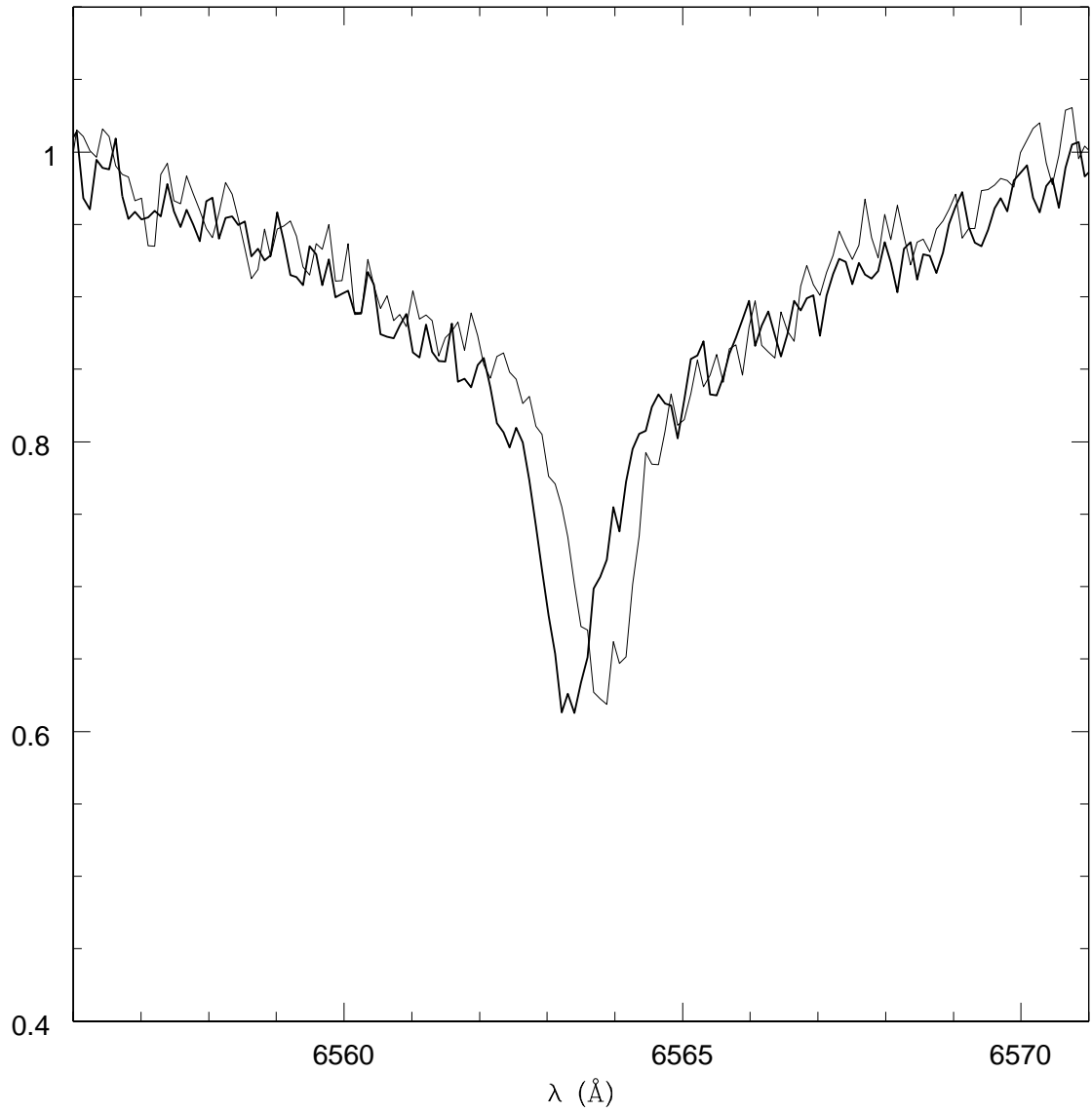}
\caption{The average of 30 blue shift spectra (thick line) and 30 red shift spectra (thin line). The spectra included in each average is measured with a red or blue shift greater than 6.9 km/s.}
\end{figure}
\begin{figure}

\figurenum{7}
\plotone{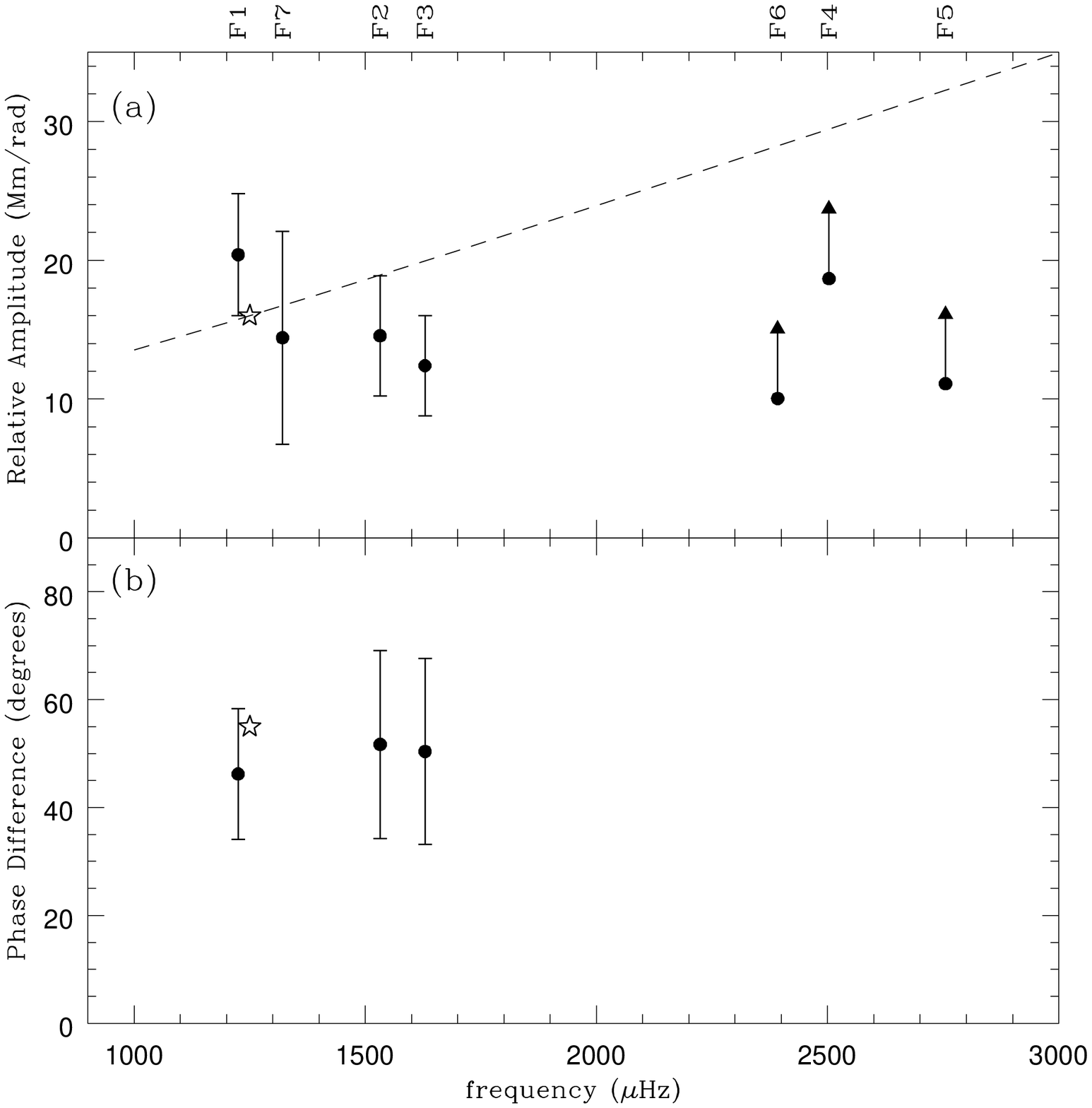}
\caption{A plot of the relative amplitude $R_v$ \textbf{(a)}, and the phase 
difference $\Delta \Phi$ \textbf{(b)} versus frequency for each mode.  
The star in each plot represents the prediction made by \citet{GW99b} with 
their nonadiabatic calculations of g-modes. The dashed line represents the 
increase in $R_v$ expected from the predictions of \citet{GW99} (for $\tau_c$ = 250~s).}
\end{figure}

Figure 7 shows our measurements of $R_v$ and $\Delta \Phi$ as a function of
frequency for the purpose of looking for these trends. Neither plot of the data appears to behave 
as expected, basically showing no obvious variations with frequency. Unfortunately, 
the comparison to the model is drastically limited by the size of our errors.
This is especially notable for modes F4, F5 and F6. Their flux amplitudes
are likely inflated by noise and thus, the values for $R_v$ are underestimated.
For this reason, we indicate lower limits for these modes in Figure 7. 
To compare the measurements with the model we have plotted the trend in
$R_v$ expected by theory in Figure 7. 
The predicted trend is not apparent in our data.
However, because of our large error bars we cannot rule it out.
We are not the first to fail to detect the trend; 
time-resolved spectra of HS~0507+0434B \citep{Ko02} failed to show the predicted
trend for $R_v$ and showed the opposite trend for $\Delta\Phi$.   

\section{The Combination Frequency in the Velocity Curve}

The analytic theory presented by \citet{GW99} makes specific predictions
about the existence and behavior of combination frequencies in the Fourier
transforms of ZZ Ceti light curves.  In their simplest models, which do not
incorporate the effects of changing convection zone thickness during
pulsation cycles, the flux variations at the photosphere are a diminished
and delayed version of the variations that occur at the base of the
convection zone.  The size of the diminution and delay for a given mode is
dependent on the thermal adjustment time of the convection zone and the mode
frequency.  As long as neither of these change, the effects are linear and
do not generate combination frequencies (assuming the
variations at the convection base are sinusoidal).  However, as first
\citet{Br83}
and later \citet{Wu01} point out, this is unrealistic.  The modes in ZZ
Cetis are large enough to substantially increase the surface temperature of
the star.  Equilibrium models of hotter ZZ Ceti stars have thinner surface
convection zones, suggesting that pulsating stars will have thinner
convection zones at pulsation maxima than at minima.  The detailed
calculations of \citet{Br92} and \citet{IK01} confirm this suggestion.
The effect of this change in convection zone thickness is that the flux is
less diminished at maximum than at minimum, i.e. the maxima rise higher
above the mean flux than the minima descend below it.  This property of the
models corresponds qualitatively, and often quantitatively, to the behavior
of the light curves we see.

In the Fourier transforms, this behavior of the models predicts combination
frequencies and harmonics that are nearly in phase with the modes that
generate them, a condition written by \citet{Vu00b} as $\phi_r =
\phi_{1,2} -
(\phi_1 + \phi_2) \approx 0$, where $\phi_{1,2}$ is the phase of the
combination and
$\phi_1$ and $\phi_2$ are the phases of its parent modes. The analysis of
modes in G~29-38
by \citet{Vu00b} shows that the combination frequencies show $\phi_r$ near an
average value of $22\degr$, consistent with their harmonic distortions
being generated by the changing convection zone depth, just as the models
describe.

The prediction of the models with respect to velocity combinations is harder
to discern.  \citet{Br90} and \citet{Wu01} assume that turbulent viscosity
in the
convection zone enforces uniform velocity with depth.  Thus their model
convection zone does not introduce any delay or diminishment of velocities
between its base and the surface, so the mechanism responsible for
the distorted light curves does not operate on the velocites.  Furthermore,
\citet{Br92}
argues that second order perturbations to the horizontal motions should be
smaller
than the linear perturbations by a factor of 10 or more.  Based on this
argument, Brickhill
\emph{imposes} sinusoidal pressure variations and displacements in his
model.  Because his model
does not allow feedback that could alter the imposed variations, it can
never exhibit
non-sinusoidal horizontal motions. The models of \citet{IK01} and the
analytic treatment
of \citet{Wu01} also assume sinusoidal pressure variations and have no
feedback. Consequently,
none of the published models have any ability to predict the shape of
velocity curves:
they only reap what they have sown.

Fortunately, the phase of our velocity combination offers a clue to its
origin. If we define the
relative phase in the same way as for the flux combinations, then we find
from Table 1 that
$\phi_{r,velocity}= \phi_{F5} - (\phi_{F1} + \phi_{F2})  = - 95\degr$.
A phase difference of $-95\degr$ suggests a simple possibility.
If the horizontal \emph{displacements} are distorted in the same way as the
flux curves
we observe, then they would generate combination frequencies with
$\phi_{r, displacement} \approx 0$.
The velocities we measure are the time derivatives of the displacements, and
the derivative
introduces a $90\degr$ phase shift in
both of the parent frequencies and in the combination, so we expect
$\phi_{r, velocity} \approx 270\degr = -90\degr$.  That is, the relative
phase of the velocity
combination frequency we have measured is close to the value we expect if
the horizontal
displacements vary in a manner similar to the flux variations.

This does not answer the question of why the measured displacements should show this behavior.   
The observations require that in the half cycle where horizontal displacements are directed 
toward a surface anti-node, material travels farther from its equilibrium position than it 
does in the half cycle where displacements are directed toward a node.   Perhaps this is 
an overlooked consequence of the strongly non-sinusoidal temperature variations at the 
pulsation anti-nodes.
Alternatively, it is possible that the uniform (with depth) horizontal velocites enforced 
by the convection zone introduce a non-sinusoidal component into the velocities as the 
thickness of the convection zone changes during a pulsation cycle.  In the models, the 
convection zone "averages" the horizontal velocites to a value somewhere between the values 
that would be present at its base and surface if turbulent viscosity did not enforce uniform 
horizontal motion (see Figure 5 of \citet{GW99b}). As the convection zone changes depth, and 
perhaps even entirely evaporates at the anti-nodes, the effect on the surface motions might 
change during a cycle, yielding a non-sinusoidal component.  However, it is not clear that 
the relative phase of this component would match the relative phase of the combination peak 
we have measured.

We can speculate further, but we are far from understanding the
origin of the velocity combination F5. First, we cannot make any
statistical arguments on the basis of one mode, so we cannot
rule out the possibility that F1, F2, and F5 represent an example of
resonant mode coupling as described by \citet{WG01} and \citet{Dz82}, 
in which case the phase relationship measured is accidental.
Second, without actual calculations extending the works of
\citet{Br83, Br92}, \citet{Wu01} and \citet{IK01}, we do not know what
detailed models will predict for the
phase of velocity combination modes.  Nonetheless, we think the speculation
we have presented represents the most fruitful immediate direction for observational
and theoretical investigations.

\section{The Spectral Profile of G~29-38}
The time-resolved, high resolution spectroscopy of G~29-38 enables us
to explore a mystery concerning the observed line profile of the 
H$\alpha$ core of ZZ Cetis.  Our data are similar to the observations
made by \citet{Ko98} to measure the rotation rate of DA white
dwarf stars except that we have a time-series of exposures of 50 s each
instead of one, hour long exposure. Thus, the average of our spectra can
confirm their rotationally broadened fit to G~29-38, while the 
short exposure times allow us to test the possibility that the line shape
is altered by the pulsational shifts.

\begin{figure}
\figurenum{8}
\plotone{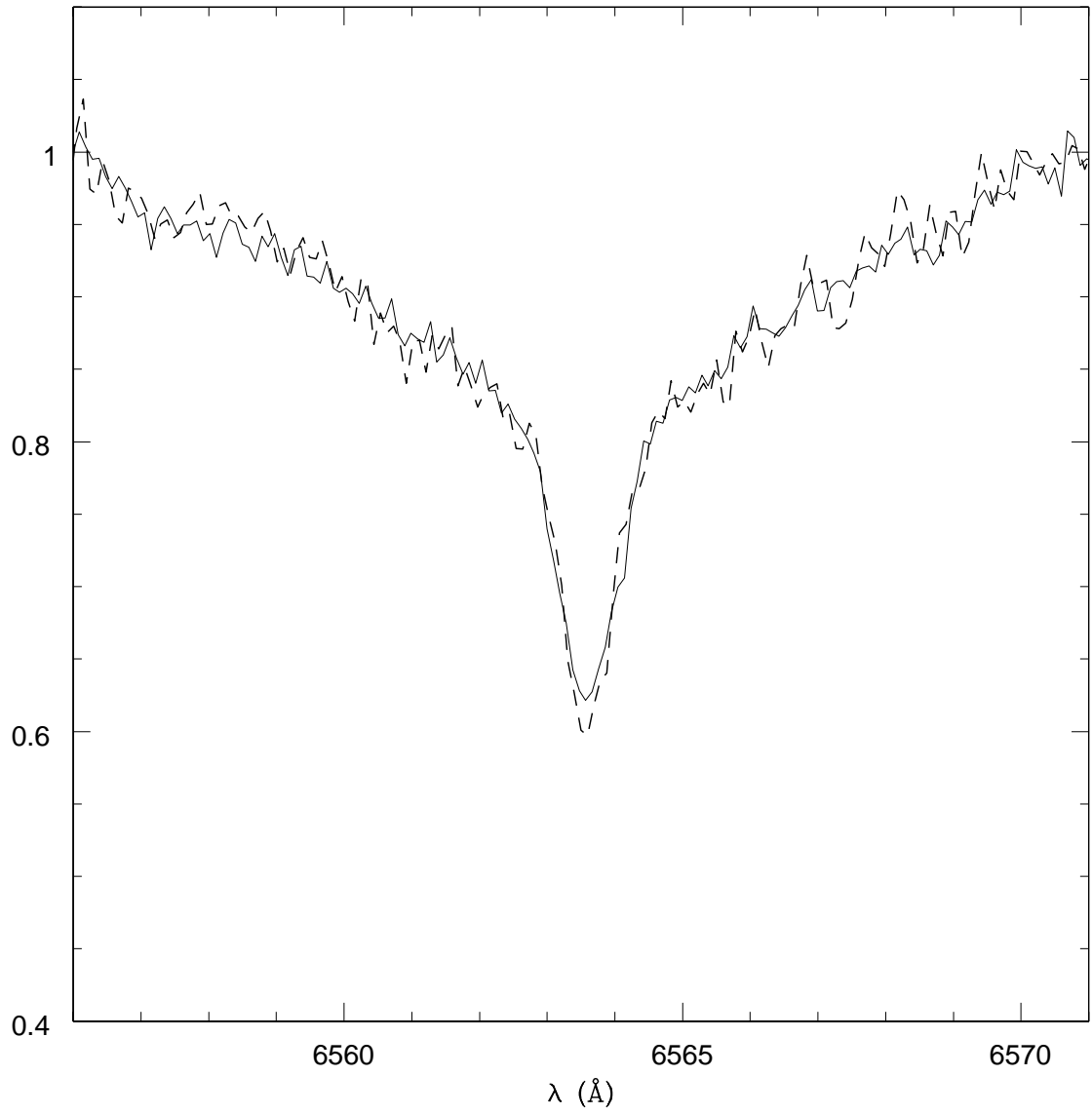}
\caption{The average of the spectra with small measured velocities (dashed line) overplotted on the average spectrum (solid line).}
\end{figure}

\begin{figure}
\figurenum{9}
\plotone{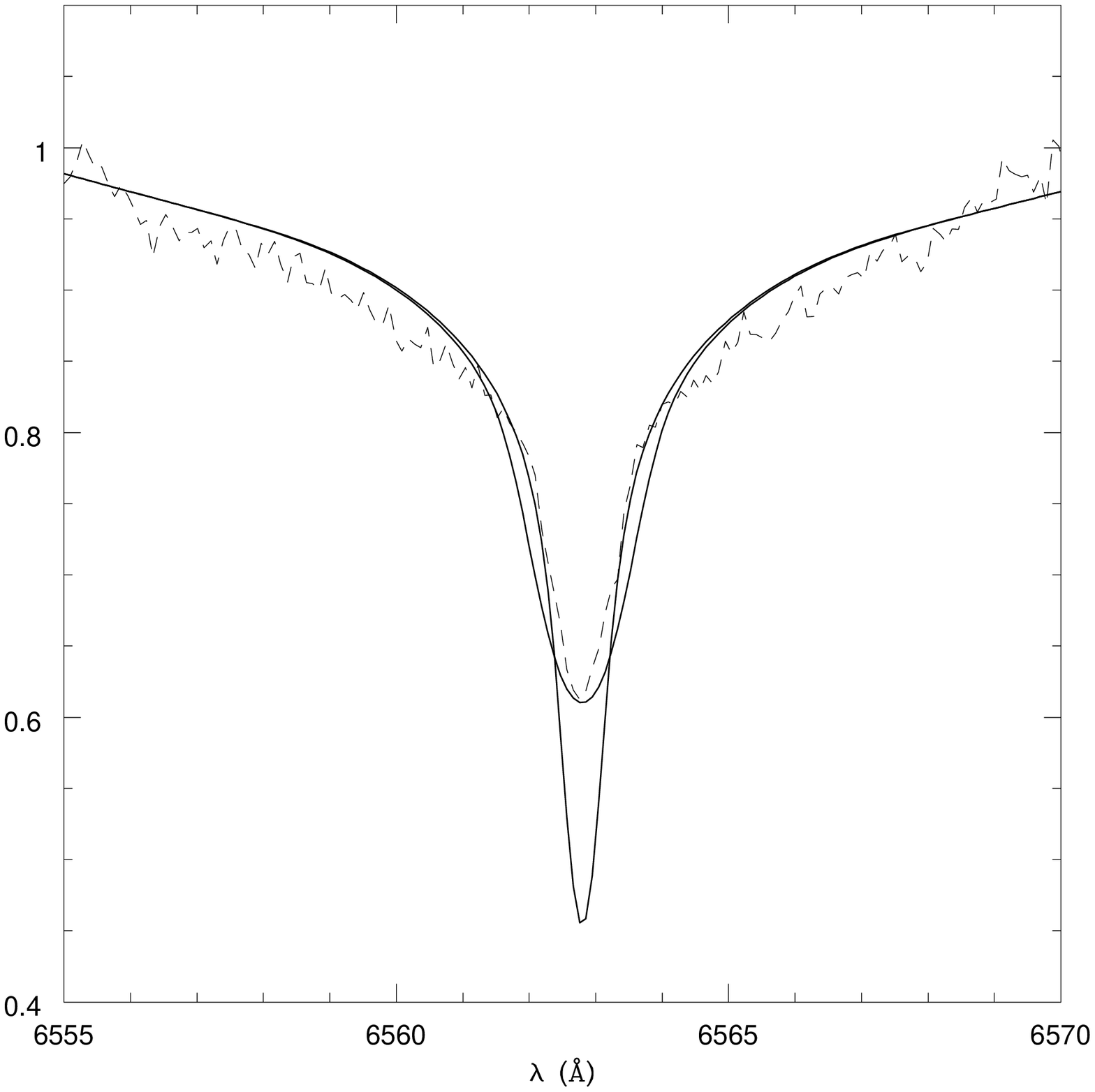}
\caption{Fits to the average spectrum. The deep core is the model with zero rotational velocity and the shallow core is the rotational broadened model corresponding to a velocity of 42.2$\pm$ 3 \kms.}
\end{figure}

\citet{Ko98} modeled the H$\alpha$ line to determine the rotation rates of 28 DA white dwarfs. 
They found that the majority of the stars have small rotation rates with an 
upper limit of 15 \kms.  However, the $v\sin i$ for each of the three ZZ Ceti stars, 
including G~29-38, is between 30 and 45 \kms.
This fitted velocity drastically exceeds what is expected from rotational splitting 
\citep{K98}, and it seems unlikely that white dwarf stars spin-up as they cool into 
the instability strip.  On close inspection of the data obtained by \citet{Ko98}, 
the shape of H$\alpha$ core
resembles the zero rotation model except for the depth of the line; 
the model is almost 20\% deeper than the spectrum.  
A large rotation model successfully fits the depth of the line, but appears
to be too broad. 
 
Reassuringly, our average spectrum is identical in shape and depth to the 1-hour exposure 
of \citet{Ko98}.  We fit the same rotationally broadened model used in \citet{Ko98} 
to our average spectrum and present the results in Figure 9.
Again, the best fit, with $ v\sin i$ of $42.2\pm 3$ \kms, is able to emulate the depth 
of the line, but does not fit the width as well as the zero rotation model.
With the increased signal-to-noise of our average spectrum,
we can confidently confirm the statement made by
\citet{Ko98} that the large rotation fit is forced by the shallow cores and
is too wide for the observations.
To quantify the discrepancy between the model and the observations 
we calculate a reduced chi-square of 3 for the fit using an
error of 1.6\% (estimated from the observational scatter outside the line core) 
for the average spectrum in Figure 8.  The odds that our rotationally broadened
model would be so poor a fit to the data by chance alone are very small. 
Accordingly, we can discard the rotational model; it is too wide
to properly fit the data.

Because rotation fails to account for the line shape, we tested the 
hypothesis that the altered shape of the spectral line is
due to pulsation.  As shown above, the g-mode pulsations create small 
Doppler shifts of the line. Since the exposure time used by \citet{Ko98} is 
much longer than the period of the modes, the spectrum could be 
smeared-out by these velocity shifts.  
If true, each of our spectra will have a line shape with
a deeper core.

Because the signal-to-noise of each individual spectrum is not 
sufficient to examine a change in the line shape, we
reduced the noise by averaging spectra with similar measured velocities.  
The high and low velocity averages, presented earlier to show the spectral 
shifts (Figure 6), have approximately the same shape as the average spectrum.   
We also averaged together 30 spectra with velocities between 1.1 and -1.1 \kms\ 
to see the line shape unaffected by the spectral shifts.
Figure 8, a plot of the small velocity and total averages, 
demonstrates how little change occurs in the line shape 
when the spectra with large velocities are excluded.  

The small velocity average shows a slightly deeper core;
however, this 2\% drop is not nearly large enough to reproduce the zero velocity
spectral line in Figure 9. By convolving the low velocity average spectrum with a gaussian
whose fwhm is equal to the root-mean-square of the velocity curve (6.8 \kms\ or 0.15 \AA), we see that
this 2\% drop is exactly what is expected for the size of the pulsation velocities.  
In fact, we would need to convolve the zero-rotation spectrum with a 
velocity four times larger than measured to achieve the average spectrum 
line depth; this convolved spectrum is also
too wide to emulate the observed spectral line.  
Though the pulsation motions do present themselves as Doppler shifts, 
these shifts do not have a significant effect on the shape of the time-averaged 
H$\alpha$\ spectrum.

Our high signal-to-noise spectra have now eliminated two possible 
explanations for the line shape of
the H$\alpha$ core of ZZ Cetis: rotation and pulsation.
Our average spectrum convincingly demonstrates that the rotationally 
broadened line does not fit the line profile.  An average spectrum
created by binning spectra with similar velocities did not result in a significantly
different line profile, eradicating the possibility that the pulsations 
truncate an intrinsically deeper line.  The only remaining explaination is that
some aspect of the stellar 
atmosphere must not be accounted for in the model spectrum. 
The shallow core suggests that the outer atmosphere is hotter than expected.  We do
not know a good physical reason for this, and therefore, must leave the question 
concerning the ZZ Ceti line shapes unanswered.

\section{Conclusions}
We have introduced high resolution, time-resolved spectroscopy to the study
of cool white dwarf pulsators with observations of the DAV G~29-38. 
This technique enabled us to measure the line-of-sight velocities 
and the flux variations associated with the g-mode pulsations. 
Despite the low signal-to-noise for each individual spectrum, the fits to the
H$\alpha$ core successfully reveal the Doppler shift of each spectrum, revealing
seven velocity modes with corresponding flux modes. 
The obvious
spectral shifts and the frequency correlation between velocity and flux modes 
confirm the previous velocity detections in G~29-38 by \citet{VK2000} 
with low-resolution spectra.

As with the low resolution data, we compared the amplitudes and phases
of the velocity and flux measurements of each mode.  To that end, we calculated the 
relative amplitude $R_v$ and phase difference $\Delta \Phi$ 
of each mode.  Overall, these values agree with the previous measurements
of G~29-38 \citep{VK2000}. The value of $\Delta \Phi$ for the three 
largest modes shows that 
the velocity maximum leads the flux maximum by less than a quarter cycle, as expected.  
The values of $R_v$ are in the same range as the previous measurements. 
None of the values are as high as the mode identified to be $\ell = 2$ \citep{C00}, 
indicating that all the modes measured here have a spherical degree of one. 

We also compared the relative amplitude and phase difference with the 
predictions made by the models of \citet{GW99} and \citet{GW99b}
which describe the pulsations on ZZ Ceti stars in the context of a 
convective driving theory.
The values of $R_v$ and $\Delta \Phi$ agree with the model;
however, they do not show the trend with frequency predicted
from the convection zone's interaction with the flux variations. 
Considering the size of our errors, this lack of agreement has
questionable value. Other observations \citep{Ko02, VK2000}
have also failed to see the predicted trends, indicating a possible
problem with the convective driving model. 

Perhaps not surprisingly, our new technique has uncovered a new phenomenon---
a velocity combination frequency.
Although it is dangerous to generalize from a single detection, it appears that the
surface velocity variations for large amplitude ZZ Cetis can experience
harmonic distortion just as the flux variations do. We have discussed this
as the product of an
underlying distortion in the horizontal displacements that has properties
similar to the
distortions in the flux variations. Unfortunately, none of the existing
models are capable of
predicting, or even reproducing, this behavior.  We believe that some effort 
in this direction could resolve many
mysteries surrounding G~29-38 and the entire class of ZZ Ceti pulsators.

Finally, we investigated
the effects of pulsation velocities on the shape of the H$\alpha$ line core.
The Doppler shifts could potentially broaden the line when observed with a long
exposure. Our short exposure times enabled us to examine the line
shape of the NLTE core of H$\alpha$ without this effect.  We discovered
that removing the spectra with large Doppler shifts from the average line 
did not significantly alter the shape of the
line, indicating that the pulsations are not causing the broadened profile. 
Rotation, another notorious line broadener, also appears unlikely because
the best rotation model failed to fit the width of the line
while yielding an unbelievably large velocity \citep{Ko98}. With both 
pulsation and rotation eliminated as potential answers, 
the cause for the truncated shape of the H$\alpha$ core
remains a mystery.

\acknowledgments
The observations for this paper were taken at the W. M. Keck Observatory, which is operated 
by the California Association for Research in Astronomy, a scientific partnership among 
the California Institute of Technology, the University of California, and the National 
Aeronautics and Space Administration.  It was made possible by the generous financial 
support of the W. M. Keck Foundation. We also recognize support from the National Science 
Foundation through grant AST-0094289 and from the DLR (Deutsches Zentrum f\"ur Luft- und Raumfahrt) through
grant 50 OR96173.  C. Clemens acknowledges the Alfred P. Sloan 
Foundation, S. Thompson would like to thank the North Carolina Space Grant 
Consortium for their support and M. H. van Kerkwijk acknowledges support by a fellowship
of the Royal Netherlands Academy of Science.

\end{document}